\documentclass[final,5p,times,twocolumn]{elsarticle}

\usepackage{amssymb}
\usepackage{lipsum}
\newcommand{\dd}{\mathrm{d}}
\newcommand{\nn}{\nonumber}
\newcommand{\hata}{\hat{a}}
\newcommand{\hatcq}{\hat{Q}}

\usepackage{graphicx} % Required for inserting images
\usepackage{geometry}
\usepackage{amsmath}
\usepackage{mathrsfs}
\usepackage{booktabs}
\usepackage{float}
\usepackage{diagbox}

\usepackage{dcolumn}% Align table columns on decimal point
\usepackage{bm}% bold math
\usepackage{wasysym}
\usepackage{xcolor}

\journal{Physics Letters B}

\begin{document}

\begin{frontmatter}

\title{Dragging of the particle spin and spin-spin coupling effect on its periapsis shift}

\author[first]{Shaofei Xu}
\affiliation[first]{
organization={School of Physics and Technology, Wuhan University},
city={Wuhan},
postcode={430072},
state={Hubei Prov.},
country={China}}

\author[corresponding]{Junji Jia\corref{cor1}}
\cortext[cor1]{Corresponding author: junjijia@whu.edu.cn}
\affiliation[corresponding]{
organization={Department of Astronomy \& MOE Key Laboratory of Artificial Micro- and Nano-structures, School of Physics and Technology, Wuhan University},
city={Wuhan},
postcode={430072},
state={Hubei Prov.},
country={China}}

\begin{abstract}
The periapsis shift (PS) of spinning test particles in the equatorial plane of arbitrary stationary and axisymmetric spacetime is studied using the post-Newtonian method. The result is expressed as a half-integer power series of $M/p$ where $M$ is the spacetime mass and $p$ is the semilatus rectum. The coefficients of the series are polynomials of the particle spin, the asymptotic expansion coefficients of the metric functions and the eccentricity of the orbit. The particle spin is shown to have a similar effect as the Lense-Thirring (LT) effect on the PS, and both of them appear from the $(M/p)^{-3/2}$ order in the PS. The coupling between the spacetime and particle spins will increase (or decrease) the PS if they are parallel (or antiparallel). For Jupiter and Saturn rotating around the Sun and exceptionally designed satellites around Mercury and Moon, the particle spin effect can be comparable to the LT one in size. The PS in other spacetime studied are not distinguishable from that in the Kerr spacetime to the $(M/p)^{-3/2}$ order.
\end{abstract}



\end{frontmatter}

%\tableofcontents



\section{Introduction}

In general relativity (GR), when a test particle rotates around a spherically symmetric massive object along a bound orbit, its periapsis will precess, which is a distinguishing feature of GR compared to Newtonian's universal gravitational law, in which the bound orbit is always elliptic.
This precession was first successfully used by Einstein to explain the extra precession of Mercury's perihelion to support the correctness of GR \cite{einstein1916}. Since the amount of the periapsis precession, i.e. the periapsis shift (PS), is related to the nature of the spacetime or the gravitational theory, nowadays such PS has been used as a tool to measure spacetime parameters \cite{binary1988,GRAVITY:2020gka} or to verify gravitational theories \cite{lucchesi2011,Everitt:2011hp}.

Among the effects of various spacetime parameters on the PS, the Lense-Thirring (LT) effect due to the spin of the spacetime is one of the most famous beside the lowest order value of $6\pi M/p$ where $M$ is the spacetime mass and $p$ is the semilatus rectum. Beside the central object however, the orbiting test particle is usually also spinning. The motion of spinning test particles in gravity is described by the Mathisson-Papapetrou-Dixon (MPD) equations \cite{Mathisson:1937zz,Papapetrou:1951pa,Dixon1964}, which are different from the geodesic equation and therefore in principle can also affect the PS.

In this work, we will study the PS
of test particles in the equatorial plane of arbitrary stationary and axisymmetric (SAS) spacetimes using the post-Newtonian (PN) method. Our motivation is to find out how the particle spin, in addition to the spacetime parameters and other kinetic variables of the orbit, will affect the PS. We are particularly interested in the relative size of the new effect comparing to previously known effects, and its potential implications for PS in astronomy.

The work is organized as follows. In Sec. \ref{sec:theo}, the MPD equation and its realization in general SAS is described. Sec. \ref{sec:pnm} applies the PN method to find the formula for the PS. This result is applied to particular spacetimes in Sec. \ref{sec:parts} and its application in astronomy is discussed in detail in Subsec. \ref{subsec:knapp}. Sec. \ref{sec:conc} ends the paper with a short conclusion and discussion. Throughout the paper, the natural units $G=c=1$ are used.

\section{Equation of motion and the PS\label{sec:theo}}

The motion of a spinning test particle in a general spacetime can be described by the MPD equations \cite{Hojman1975}
    \begin{align}
        \frac{\mathrm{D}p^{\mu}}{\mathrm{D}\lambda}=&-\frac{1}{2}R^{\mu}_{\nu\rho\sigma}u^{\nu}S^{\rho\sigma},\label{eq:mpdp}\\
        \frac{\mathrm{D}S^{\mu\nu}}{\mathrm{D}\lambda}=&2p^{[\mu}u^{\nu]},\label{eq:mpds}
    \end{align}
where $\mathrm{D}$ stands for the total derivative, $p^{\mu},\,u^{\mu}\equiv \dd x^{\mu}/\dd\lambda$ and $S^{\mu\nu}$ are the generalized total momentum, four-velocity and antisymmetric spin tensor of the test particle respectively. Since the number of free functions in these equations is more than that of the equations, we will supplement this with the Tulczyjew constraint \cite{Tulczyjew}
\begin{equation}
    S^{\mu\nu}p_{\nu}=0.
    \label{equ:Tulczyjew}
\end{equation}
Moreover, the momentum $p^\mu$ and spin $S_{\mu\nu}$ satisfy the normalization conditions
    \begin{align}
        p^{\mu}p_{\mu}=&-m^2 ,\label{eq:consm} \\
        \frac{1}{2}S^{\mu\nu}S_{\mu\nu}=&J_m^2, \label{eq:consjm}
    \end{align}
where $m$ and $J_m$ are the mass and spin angular momentum of the test particle, respectively. From the spin tensor $S^{\mu\nu}$, we can also define a vectorial spin
\begin{align}
    S_\mu=\frac{\sqrt{-g}}{2m}\epsilon_{\mu\alpha\beta\gamma}S^{\alpha\beta}p^\gamma,
    \label{equ:Smu}
\end{align}
where $g=\mathrm{det}(g_{\mu\nu})$ and $\epsilon$ is the Levi-Civita tensor.

In this work, we will study the motion of such test particles in an arbitrary SAS spacetime whose line element can always be written as
\begin{equation}
    \dd s^2=-A(r,\theta)\dd t^2+B(r,\theta)\dd t\dd\varphi+D(r,\theta)\dd r^2+C(r,\theta)\dd\varphi^2,
    \label{equ:spacetime}
\end{equation}
where $(t,\,r,\,\theta,\,\varphi)$ are the coordinates and $A,\,B,\,C,\,D$ are functions of $r$ and $\theta$ only. The symmetry of the spacetime allows two conserved quantities, the specific energy $E$ and angular momentum $L$ of the test particle. For simplicity, we will concentrate on motions in the equatorial plane $(\theta(\tau)=\pi/2)$ and satisfying $S_t=S_r=S_\varphi=0$ and $S_\theta\neq 0$. This kind of motion can be shown to exist if we choose $S^{\mu\theta}=0=p^\theta$ and the metric function satisfies $\partial_\theta A=\partial_\theta B=\partial_\theta C=\partial_\theta D=0$ on the equatorial plane \cite{Zhang:2022rnn}.

Under these conditions, then one can use Eqs. \eqref{equ:Tulczyjew} and \eqref{eq:consjm} to express $(S^{tr},\,S^{t\varphi},\,S^{r\varphi})$ in terms of $(p^t,\,p^r,\,p^\varphi)$ as well as the metric functions. Further using the conserved $E$ and $L$ as well as the condition \eqref{eq:consm}, the momentum $(p^t,\,p^r,\,p^\varphi)$ can be expressed using the metric functions and $(E,\,L)$ only. Finally using the MPD equations \eqref{eq:mpdp}, the equations of motion are found in the form of $\dd \varphi/\dd r$ and $\dd t/\dd r$. We refer the readers to Ref. \cite{Zhang:2022rnn} and references therein for details of this process. The ratio between these two quantities provides the relation needed in this work
\begin{align}
    \frac{\dd\varphi}{\dd r}=&\frac{1}{\eta_3^2(B^2+4AC)D\sqrt{(p^{r})^2}}\bigg(\eta_1\eta_7\nonumber\\
    &+8\eta_2\left\{[-16BD-8D\alpha^\prime (B^2+4AC)]+\alpha\eta_6+\alpha^2\eta_5\right\}\bigg)
    \label{eq:dphiodr}
\end{align}
where
\begin{subequations}
\begin{align}
(p^{r})^2=&\frac{1}{D}\left[\frac{64(-A\eta_1^2+B\eta_1\eta_2+C\eta_2^2)}{(B^2+4AC)\eta_3^2}-m^2\right], \label{equ:Ptphir}\\
\eta_1=&4L+\alpha(B^\prime L+2C^\prime E),\label{eq:eta1def}\\
\eta_2=&4E+\alpha(2A^\prime L-B^\prime E),\label{eq:eta2def}\\
\eta_3=&-16+\alpha^2(B^{\prime 2}+4A^\prime C^\prime ),\label{eq:eta3def}\\
\eta_5=&BD(B^{\prime 2}+4A^\prime C^\prime )-DB^\prime (B^2+4AC)^\prime \nonumber\\
&-(B^2+4AC)(B^\prime D^\prime -2D^{\prime \prime}),\\
\eta_6=&-4[D(B^2+4AC)]^\prime,\\
\eta_7=&256AD+8\alpha^2[2D(4CA^{\prime 2}+2BA^\prime B^\prime -AB^{\prime 2})\nonumber\\
&+2(B^2+4AC)(A^\prime D^\prime -2DA^{\prime\prime})],\\
\alpha=&-\frac{a_m}{\sqrt{D(B^2/4+AC)}},
    \label{equ:alpha}
\end{align}
\end{subequations}
and $^\prime$ denotes derivative with respect to  $r$, and $a_m=J_m/m$
is the specific spin angular momentum per mass of the test particle. In Eq. \eqref{equ:alpha} we chose the negative sign because from Eq. \eqref{equ:Smu} we have the relation
\begin{equation}
    S_{\theta}=\frac{\sqrt{-g}}{2m}\varepsilon_{\theta\alpha\beta\gamma}S^{\alpha\beta}p^{\gamma}=\sqrt{-g}m\alpha=-J_m
\end{equation}
so that when $S_\theta$ is positive (downward) the particle spin along $z$ axis is negative.

Note the effect of the particle spin is completely reflected through $\alpha$, which is proportional to $ a_m$. When $\alpha=0$, Eq. \eqref{eq:dphiodr} reduces to the geodesic equation of the spinless test particle in the given spacetime.

\section{The post-Newtonian precession\label{sec:pnm}}

To consider the PS of the spinning test particle, its orbit will have to be bounded in the radial direction. The radial boundaries $r_1$ and $r_2$ should satisfy the condition
\begin{equation}
    \frac{\dd r}{\dd\varphi} \bigg|_{r_1}=0=\frac{\dd r}{\dd\varphi} \bigg|_{r_2}.
\end{equation}
Without losing any generality, we will assume $r_1<r_2$ and therefore $r_1$ and $r_2$ are the radii of the apoapsis and periapsis
respectively.
According to Eq. \eqref{eq:dphiodr} and \eqref{equ:Ptphir}, then this is equivalent to
\begin{align}
(p^{r})^2|_{r_{1,2}}=\frac{1}{D}\left[\frac{64(-A\eta_1^2+B\eta_1\eta_2+C\eta_2^2)}{(B^2+4AC)\eta_3^2}-m^2\right]\bigg|_{r_{1,2}}=0.
    \label{equ:enLfunction}
\end{align}
It is seen from the Eqs. \eqref{eq:eta1def}-\eqref{eq:eta2def}, that both $\eta_1$ and $\eta_2$ are linear combinations of $(E,L)$ and therefore the boundary condition \eqref{equ:enLfunction} will allow us to solve $(E,L)$ as functions of $r_{1,2}$ easily. On the other hand, it is not difficult to learn that once the kinetic parameters $(E,L)$ are fixed, in principle the boundary values $r_{1,2}$ are also determined for any specific spacetime.

The PS of the orbit, then can be defined as
%we will obtain four group of $E(r_1,r_2),L(r_1,r_2)$. And we define $E>0$ and $L>0(L<0)$ represent anticlockwise (clockwise)\\
\begin{equation}
\delta=2\int_{r_1}^{r_2}\frac{\dd\varphi}{\dd r}\dd r-2\pi, \label{eq:psdef}
\end{equation}
where $\frac{\dd \varphi}{\dd r}$ is given by Eq. \eqref{eq:dphiodr}.
In general however, the integral in Eq. \eqref{eq:psdef} can not be carried out to obtain a closed form. Therefore some kinds of approximation have to be used. In Ref. \cite{He:2023joa}, we developed a near-circular approximation and two variants of the PN approximation. Here for simplicity and easier physical interpretation of the result, we will only use one of the PN methods.

To use this method, we assume that the orbit size is large and therefore the orbit is close to an ellipse, i.e., the orbit is described by the equation
\begin{equation}
\frac{1}{r}=\frac{1-e\cos\psi(\varphi)}{p},
    \label{equ:upsi}
\end{equation}
where the function $\psi(\varphi)$ characterizes how far the orbit deviates from the ellipse, which corresponds to the case $\psi(\varphi)=\varphi$. $e$ and $p$ are respectively the effective eccentricity and semilatus rectum. The PN approximation in this problem is equivalent to the large $p$ approximation. Besides, Eq. \eqref{equ:upsi} implies that the apoapsis and periapsis are at $\psi=\pi$ and $\psi=0$ respectively. The parameters $(e,\,p)$ then can be connected to $r_{1,2}$ through
\begin{align}
    \frac{1}{r_1}=\frac{1+e}{p},~\frac{1}{r_2}=\frac{1-e}{p}. \label{eq:r1r2inep}
\end{align}
Using Eq. \eqref{equ:enLfunction}, therefore the relation between $(e,\,p)$ and $(E,\,L)$ can be established.

We then can do a change of variables in PS \eqref{eq:psdef} from $r$ to $\psi$ by working out from Eq. \eqref{equ:upsi} the identity
\begin{align}
    \frac{\dd\varphi}{\dd\psi}=\frac{-er^2}{p}\sin\psi\frac{\dd\varphi}{\dd r}.
\end{align}
Substituting into Eq. \eqref{eq:psdef}, we have
\begin{align}
    \delta=&2\int_0^\pi\frac{\dd\varphi}{\dd\psi}\dd\psi-2\pi\nonumber\\
    =&2\int_0^\pi \frac{-er^2}{p}\sin\psi\frac{\dd\varphi}{\dd r}\dd\psi-2\pi
\end{align}
where $\frac{\dd\varphi}{\dd r}$ is still given by Eq. \eqref{eq:dphiodr} and all $r$ in this expression should be further substituted by $\psi$ using Eq. \eqref{equ:upsi}. After this, the PS becomes
\begin{align}
    \delta=&2\int_0^\pi F(\cos\psi;p,e,E,L)\dd\psi -2\pi,
\end{align}
where $F$ denotes the entire integrand. Let us emphasize that once the metric functions are known, then from Eq. \eqref{eq:dphiodr}, $F$ is completely determined.

Under the PN approximation, we can do a large $p$ expansion of $F(\cos\psi;p,e,E,L)$ and prove that the result can always be integrated. We refer the readers to Ref. \cite{He:2023joa} for details of this proof. Then the result after integration can also be expressed as a power series of $p^{-1}$
\begin{align}
\delta=\sum_{n=1}\frac{i_n}{p^n}
    \label{eq:psinpeel}
\end{align}
where the coefficients $i_n$ are related to the coefficients of the asymptotic expansions of the metric functions
\begin{subequations}
\begin{align}
&A(r)=\sum_{n=0}\frac{a_n}{r^n},\quad B(r)=\sum_{n=1}\frac{b_n}{r^n},\\
&C(r)=\sum_{n=0}\frac{c_n}{r^{n-2}},\quad D(r)=\sum_{n=0}\frac{d_n}{r^n}.
\end{align} \label{eq:abcdexp}
\end{subequations}
Due to the asymptotic flatness of the spacetime, here we will set $a_0=d_0=c_0=1$
and assume that $-a_1$, which is interpreted as the ADM mass in most cases, is not zero.
The first few such $i_n$'s are found to be
\begin{subequations}
    \begin{align}
        &i_0=2\frac{\pi}{\sqrt{-f_2}}-2\pi,\\
        &i_1=-\frac{3\pi f_3}{(-f_2)^{\frac{3}{2}}},\\
        &i_2=\frac{3\pi[(18+e^2)f_3^2-4(4+e^2)f_2f_4]}{8(-f_2)^{\frac{5}{2}}},
\end{align}
\end{subequations}
where
    \begin{align}
            f_2=&-1+\frac{E^2-m^2}{4(L-E a_m)^2}\left\{4[(c_1-d_1)^2+2c_2-d_2]\right.\nonumber\\
            &-\frac{E}{L-E a_m}\left[4 (b_1 (a_1+d_1-2 c_1)-b_2)\right.\nonumber\\
            &+ a_m \big(a_1 (8 c_1-6 d_1-3a_1)+(c_1-7d_1)(d_1-c_1)\nonumber\\
            &\left.\left.+4 a_2+4 c_2+4 d_2\big)\right]+\frac{3E^2(b_1- a_m(a_1+d_1))^2}{(L-E a_m)^2}\right\}\nonumber\\
            &+\frac{E^2}{4(L-E a_m)^2}\left\{a_1(a_1-2c_1+d_1)-a_2\right.\nonumber\\
            &\left.+\frac{4E}{L-E a_m}(-a_1b_2+ a_m(a_1^2+a_1d_1))\right\},
\end{align}
and $f_3$ and $f_4$, although simple to obtain, are too lengthy to present here.

The result \eqref{eq:psinpeel} however, is still not a pure large $p$ expansion due to the presence of kinetic variables $(E,\,L)$ which are equivalent to $(p,\,e)$. As stated after Eq. \eqref{eq:r1r2inep}, using Eq. \eqref{equ:enLfunction}
\begin{equation}
\frac{64(-A\eta_1^2+B\eta_1\eta_2+C\eta_2^2)}{(B^2+4AC)\eta_3^2}-m^2\bigg|_{r=\frac{p}{1\pm e}}=0,
\end{equation}
and substituting in the expansions \eqref{eq:abcdexp} and carrying out the large $p$ expansion too, the $(E,\,L)$ can be linked to $(p,\,e)$ as
\begin{equation}
    E=\sum_{n=0}^{\infty}\frac{e_n}{p^{n/2}},\quad L=\sum_{n=-1}^{\infty}\frac{l_n}{p^{n/2}},
\end{equation}
where the first few coefficients $e_n,\, l_n$ are
\begin{subequations}
    \begin{align}
        &e_0=m,~e_1=e_3=0,~e_2=\frac{(1-e^2)ma_1}{4},\\
        &e_4=\frac{(1-e^2)^2ma_1(3a_1-4c_1)}{32},\\
        &e_5=\frac{(1-e^2)^2m\sqrt{-a_1}(b_1+ a_md_1)}{4\sqrt{2}},\\
        &l_{-1}=s_Lm\sqrt{\frac{-a_1}{2}},~l_0=m a_m,\\
        &l_1=s_L\frac{(3+e^2)m(a_1-c_1)a_1-4ma_2}{4\sqrt{-2a_1}},
    \end{align}
\end{subequations}
where in $l_{-1,1}$ the $s_L=\textrm{sign}(L)=\pm1$ is the sign of $L$.
Note that the particle spin $a_m$ does not manifest in $E$ until the $p^{-5/2}$ order, while it shows up in $L$ from the $p^0$ order, which suggests the main influence of the particle spin on the PS occurs through its coupling with the orbital angular momentum.
Finally, plugging $(E,\,L)$ into Eq. \eqref{eq:psinpeel}, the PS in the form of power series of $p$ is found as
\begin{align}    \delta_{\mathrm{PN}}=\sum_{n=2}^\infty\frac{s_L^n\delta_{n/2}}{(-a_1)^{n/2}p^{n/2}}
\label{eq:psf}
\end{align}
where $-a_1$ provides a basic length scale against which $p$ can be measured. The coefficients $\delta_{n/2}$ of the $p^{-n/2}~(n=2,3,\cdots)$ order contribution, are given by
\begin{subequations}
\label{eq:deltan2s}
\begin{align}
\delta_1=&-\pi \left[( c_1+  d_1)a_1-2
    a_1^2+2  a_2\right],\label{eq:d1exp}\\
\delta_{3/2}=&-\sqrt{2}\pi [( a_1-2d_1)a_1  a_m+2 a_1b_1],\\
\delta_{2}=&\frac{\pi}{8}  \bigg\{3a_1 b_1  a_m+40 a_1^4+8a_1^3 [c_1 \left(e^2-4\right)-d_1]\nonumber\\
   &-a_1^2 [80 a_2+2 c_1 d_1 \left(e^2-2\right)-\left(4e^2+8\right)c_2\nonumber\\
   &+c_1^2 \left(5 e^2+2\right)+(e^2+2)\left(d_1^2-4d_2\right)]\nonumber\\
   &\left.+8 a_1 [a_2 \left(4c_1-e^2c_1+d_1\right)+6 a_3]-8a_2^2\right\},\\
   \delta_{5/2}=&\frac{\sqrt{2}\pi a_1}{4}\left\{\left[-3\left(7+2e^2\right)a_1^3+\left(\left(11+e^2\right)c_1+4\left(3-e^2\right)d_1\right)a_1^2\right.\right.\nonumber\\
&+\left(4\left(7+2e^2\right)a_2+\left(6e^2-10\right)c_1d_1-4\left(5+e^2\right)c_2\right.\nn\\
   &\left.+\left(e^2-20\right)d_2+(e^2+5)c_1^2+(3e^2+11)d_1^2\right)a_1 \nn\\
   &\left.-4\left(3-e^2\right)d_1a_2
   \right]a_m+ \left(e^2+5\right)4 a_1 b_2\nn\\
&\left.+ \left[4  \left(3-e^2\right)a_2 -32 a_1^2+(10-6e^2)
  a_1c_1+4a_1d_1\right]b_1\right\}.
    \end{align}
\end{subequations}
The formula \eqref{eq:psf} together with \eqref{eq:deltan2s} are one of the key results of the work. It shows exclusively how the particle spin, spacetime parameters as well as kinetic variables affect the PS.

Since in most physically relevant spacetimes, $-a_1$ and $d_1$, $a_2$ and $-b_1/4$ are interpreted respectively as the mass $M$, charge square $Q^2$ and spin angular momentum $J_M$ of the spacetime, we see from Eq. \eqref{eq:d1exp} that in general only $M$ and $Q$ of the spacetime appear in the leading order (order $p^{-1}$) of the PS. While the spacetime spin $J_M$ or equivalently the specific spin angular momentum $a_M\equiv J_M/M$, and particle spin $ a_m$ both start appearing only from the next order (order $p^{-3/2}$). Moreover, because of the sign $s_L$, the contribution of $a_M$ and $ a_m$ at their leading order is proportional to their relative signs to the orbital angular momentum. In other words, when they are parallel to $L$, the PS is decreased, and otherwise, enhanced. Also from this order, we see that if the central object does not have any spin ($b_1=0$), then the particle spin $a_m$ can create a similar frame-dragging effect to the PS as the LT effect. Furthermore, it is also noticeable that if the PS is considered to one more order higher ($p^{-2}$ order), the coupling term between the two spins $b_1$ (or $a_M$) and $ a_m$ will appear. These features are more clearly seen from the KN result in Eq. \eqref{eq:knjmps}.

\section{Results in particular spacetimes\label{sec:parts}}

To see more directly the effect of various spacetime and particle parameters on this PS, in this section, we apply the PS \eqref{eq:psf} to the KN spacetime.

\subsection{KN spacetime result and implications\label{subsec:knapp}}
The metric functions of the KN spacetimes in the form of Eq. \eqref{equ:spacetime} are known as
\begin{equation}
    \begin{aligned}
        A(r,\theta)=&1-\frac{2Mr-Q^2}{\Sigma},\quad B(r,\theta)=-\frac{2\sin^2\theta  a_M(2Mr-Q^2)}{\Sigma},\\
        C(r,\theta)=&\frac{(r^2+  a_M^2)^2-\Delta  a_M^2}{\Sigma}\sin^2\theta,\quad D(r,\theta)=\frac{\Sigma}{\Delta},
    \end{aligned}
    \label{equ:KNspacetime}
\end{equation}
where $M,\,Q,\, a_M$ are the mass, charge and specific spin angular momentum of the spacetime respectively, and
\begin{equation}
\Sigma=r^2+ a_M^2\cos^2\theta,\quad\Delta=r^2-2Mr+ a_M^2+Q^2.
\end{equation}
Their asymptotic expansions in the equatorial plane are
\begin{subequations}
    \begin{align}
        A(r)=&1-\frac{2M}{r}+\frac{Q^2}{r^2}+\mathcal{O}(r)^{-3},\\
        B(r)=&-\frac{4 a_MM}{r}+\frac{2 a_MQ^2}{r^2}+\mathcal{O}(r)^{-3},\\
        C(r)=&r^2+ a_M^2+\mathcal{O}(r)^{-1},\\
        D(r)=&1+\frac{2M}{r}+\frac{4M^2- a_M^2-Q^2}{r^2}+\mathcal{O}(r)^{-3},
    \end{align}
\end{subequations}
Substituting these coefficients into Eq. \eqref{eq:psf}, the PS in the KN spacetime is found to be
\begin{align}
\delta_{\mathrm{KN}}= &\pi (6-\hatcq^2)\frac{M}{p}-2 \pi s_L(4 \hata_M+3
\hata_m)\left(\frac{M}{p}\right)^{3/2}\nn\\
&+\frac{\pi}{4 }\left[6
   (e^2+18)-2 (e^2+24)
\hatcq^2-\hatcq^4\right.\nn\\
&\left.+ 12  \hata_M^2+24  \hata_M  \hata_m)\right]\left(\frac{M}{p}\right)^2 \nn\\
&+ \pi s_L \left\{
\left[-3
  (e^2+14) +(2e^2+15)   \hatcq^2\right]\hata_m\right.\nn\\
&\left. -8  (9 -2 \hatcq^2)\hata_M\right\}\left(\frac{M}{p}\right)^{5/2}\nn\\
&+\frac{\pi}{8}\left\{
-180\left(6+e^2\right)+6\left(126+19e^2\right)\hatcq^2\right.\nn\\
&-2(31-e^2)\hatcq^4+\hatcq^6+
12\left(3e^2-4\right)\hata_m^2\nn\\
&-8\left[3\left(38+3e^2\right)-2\left(6+e^2\right)\hatcq^2\right]\hata_M\hata_m\nn\\
&\left.+\left[24\left(3e^2-25\right)+4\left(11-e^2\right)\hatcq^2\right]\hata_M^2\right\}\left(\frac{M}{p}\right)^3\nn\\
&+\mathcal{O}(M/p)^{-7/2},\label{eq:knjmps}
\end{align}
where $\hat{x}\equiv x/M$ for all hatted quantities. In the first order that it appears, i.e., the $(M/p)^{3/2}$ order, the $\hata_M$ term corresponds to the famous LT effect on the PS. This effect is more conventionally expressed using the rate of precession $\dot{\varpi}_\mathrm{LT}$ obtainable by substituting $p=a(1-e^2)$ where $a$ is the semimajor axis into this term and then dividing the period $T=2\pi (a^3/M)^{1/2}$
\begin{align}
\dot{\varpi}_\mathrm{LT}=-\frac{4J_M}{a^3(1-e^2)^{3/2}}.
\end{align}
It is also seen that the particle spin $\hata_m$ starts appearing from the same $(M/p)^{3/2}$ order. For spinless particles, setting $\hata_m$ to zero,  this PS reduces to Eq. (77) of Ref. \cite{He:2023joa}. Setting $e=0$ and $Q=0$, this reduces to the results in Ref.  \cite{Mukherjee:2018zug,Hackmann:2014tga} which studied the PS of a circular orbit in Kerr spacetime. We also remind that the PS \eqref{eq:knjmps} works not only for KN black hole (BH) spacetime but also naked singularities when $\hata_M>1$.

At its leading order, the contribution of $\hata_m$ to the PS depends linearly on $\hata_m$
and is sensitive to the sign of $s_L$. When the spin angular momentum of the test particle is parallel (or antiparallel) to its orbital angular momentum, the PS is decreased (or increased) by the particle spin. This is consistent with the observation that parallel (or antiparallel) $\hata_m$ and $L$ will decrease (or increase) the bending of the trajectory in Ref. \cite{Zhang:2022rnn}. While from the $(M/p)^2$ order, the nonlinear contributions $\hata_M^2$ and the spin-spin coupling term $\hata_m\hata_M$ start to appear. The former always increases while the latter increases (or decreases) the PS when the two spins are parallel (or antiparallel).
Interestingly, if we think of the effect of the coupling between $\hata_m$ and $\hata_M$ as an extra force of GR (beyond the Newtonian theory), then the sign of this force agrees with the Ampere's force between two current-carrying wires. That is, when the spins/currents are parallel (or antiparallel), there will be an effective attractive (or repulsive) force, such that the trajectory is bent more (or less) and the PS is increased (or decreased). Finally, also note that unlike $\hata_M$, the non-linearity of $\hata_m$ appears from order $p^{-3}$.

\begin{figure}[htp!]
\centering
\includegraphics[width=0.45\textwidth]{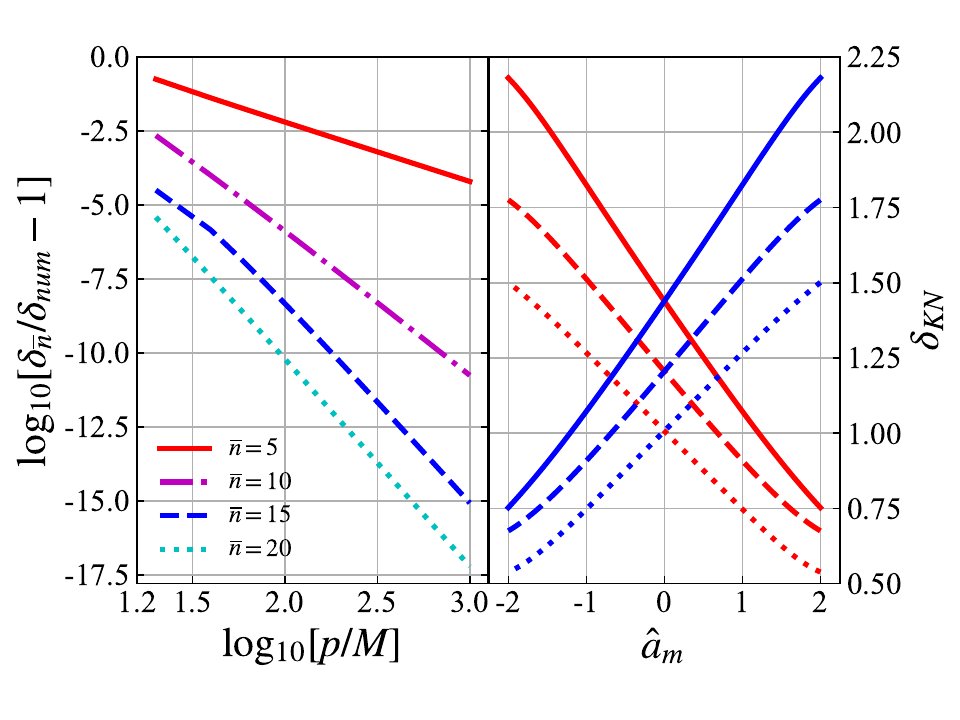}
\caption{Left: Difference between truncated PN PS \eqref{eq:knjmps} to order $\bar{n}$ and numerical PS for $p$ from $20M$ to $2000M$. $\hata_m=2,\,\hata_M=1/2$ are used. Right: $\delta_{\mathrm{KN}}$ as a function of $\hata_m$ for $s_L=+1$ (red) and $s_L=-1$ (blue) and $\hata_M=1/2~\text{(solid)},\,0~\text{(dashed)},\,-1/2~\text{(dotted)}$. $p=20M$ is used. In both plots,
$\hatcq=1/4,\,e=1/3$ are set.}
\label{fig:knplot}
\end{figure}

To illustrate these features, and more importantly show the correctness of the PS \eqref{eq:knjmps}, in
Fig. \ref{fig:knplot} (left) we plot the difference between the PS \eqref{eq:knjmps} truncated to order $\bar{n}$ and the numerical PS as a function of $p$. The latter is computed to very high accuracy and therefore can be considered as the true PS. As $\bar{n}$ increases, one observes that the PS approaches the numerical result exponentially, which shows the correctness of the PS \eqref{eq:knjmps}. In Fig. \ref{fig:knplot} (right) the PS $\delta_{\mathrm{KN}}$ itself is plotted as a function of the particle spin $\hata_m$ for several choices of $\hata_M$ and two $s_L$'s using Eq. \eqref{eq:knjmps} to high order. The effects of spacetime spin $\hata_M$ and charge $\hatcq$ have already been investigated in Ref. \cite{He:2023joa}.
It is clear that the features seen from both the $p^{-3/2},\, p^{-2},\,p^{-5/2}$ orders and higher orders are present: when $\hata_m$ is relatively small ($|\hata_m|\lesssim 1$) in this parameter setting, $\delta_{\mathrm{KN}}$ depends on $\hata_m$ linearly with the slope sign $s_L$; while when $\hata_m$ is larger ($|a_m|\gtrsim 1$), some non-linearity appears, also as dictated by the high orders of Eq. \eqref{eq:knjmps}.

\begin{figure}[htp!]
\centering
\includegraphics[width=0.23\textwidth]{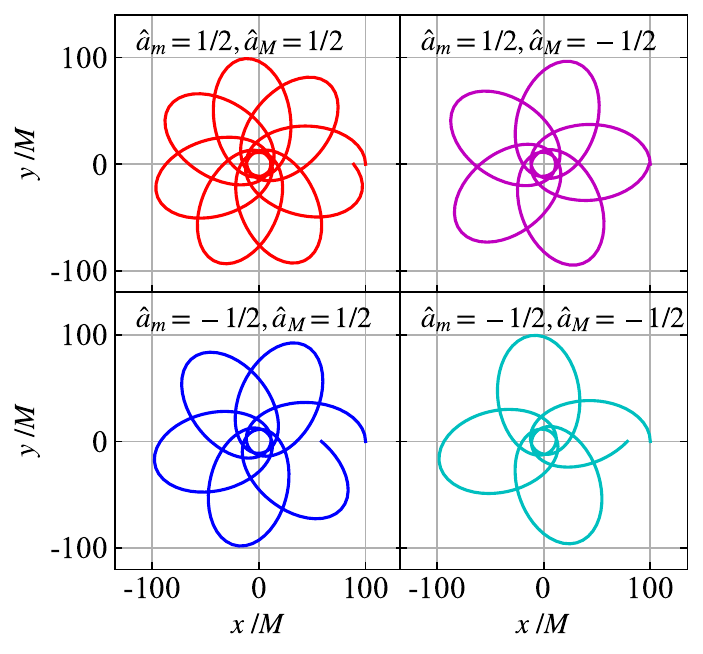}
\includegraphics[width=0.23\textwidth]{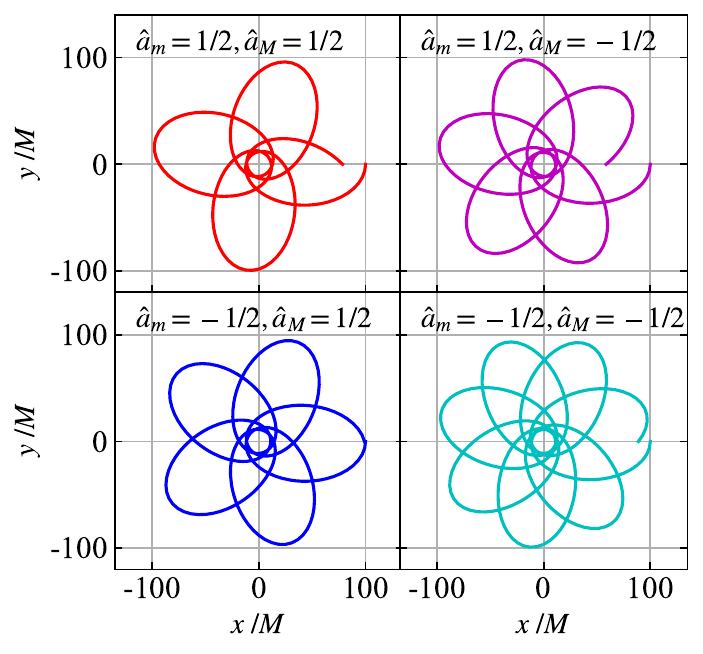}
\caption{Counterclockwise (left) and clockwise (right) rotating trajectories for $\hata_m=\pm 1/2,~\hata_M=\pm1/2$. $\hatcq=1/3,\,e=4/5,\,p=20M$ are used in both plots. All trajectories start from the positive $x$ axis.}
\label{fig:knt}
\end{figure}

The qualitative feature above, especially the dependence of $\delta_{\mathrm{KN}}$ on $\hata_m$, $\hata_M$ and the relative signs between them and $L$ are also verified in the trajectories in Fig. \ref{fig:knt}. The left and right plots correspond to the trajectories that rotate counterclockwise ($s_L=+1$) and clockwise ($s_L=-1$) respectively. It is seen from both plots that, when $\mathrm{sign}(\hata_M)=\mathrm{sign}(\hata_m)=s_L$ the PS is decreased most and when $\mathrm{sign}(\hata_M)=\mathrm{sign}(\hata_m)=-s_L$ the PS is increase most. The PS for $\mathrm{sign}(\hata_M)=-\mathrm{sign}(\hata_m)$ lie in between. Again, these are in perfect agreement with Eq. \eqref{eq:knjmps}.

Because of the importance of the LT effect in PS observations of the planets or experiments using satellites, and the fact that the $\hata_m$ term is at the same order, it becomes necessary to examine carefully in Eq. \eqref{eq:knjmps} the size of the $\hata_m$ term comparing to the $\hata_M$ term. For the eight planets orbiting the Sun, it is not hard to find using their values of spin angular momenta \cite{ Discrepant estimates of moments,Global Earth Physics,Iorio:2008mx,Mars high resolution gravity,cang:2016, Park:2017zgd,mili:2023} the ratios $\hata_m/\hata_M$ for each pairs. From Mercury to Neptune, they are
\begin{align}
&\frac{\hata_{\uparrow \mercury}}{\hata_{\uparrow \odot \mercury}}\simeq 3.1\times 10^{-5},\,
\frac{\hata_{\uparrow \venus}}{\hata_{\uparrow \odot \venus}}\simeq -4.6\times 10^{-5},\,
\frac{\hata_{\uparrow \earth}}{\hata_{\uparrow \odot \earth}}\simeq 1.1\times 10^{-2},\,\nn\\
&\frac{\hata_{\uparrow \mars}}{\hata_{\uparrow \odot \mars}}\simeq 3.1\times 10^{-3},\,
\frac{\hata_{\uparrow \jupiter}}{\hata_{\uparrow \odot \jupiter}}\simeq 2.5\times 10^{0},\,
\frac{\hata_{\uparrow \saturn}}{\hata_{\uparrow \odot \saturn}}\simeq 1.2\times 10^{0},\,\nn\\
&\frac{\hata_{\uparrow \uranus}}{\hata_{\uparrow \odot \uranus}}\simeq -3.8\times 10^{-2},\,
\frac{\hata_{\uparrow \neptune}}{\hata_{\uparrow \odot \neptune}}\simeq 2.5\times 10^{-1},\,
\end{align}
where the $\uparrow$ means that we also projected all spins to be perpendicular to the orbital planes of the corresponding planet.
It is clear that the Mercury spin is too small to affect the LT effect measurement from its PS observation. On the contrary, for Jupiter and Saturn, their own spin contributions are larger than the LT effect term due to solar rotation, although currently their observed PS still can not resolve any of these effects \cite{azps2016}. For other planets, the situations lie in between. For the planet-moon systems, we only calculated the ratio for the Earth and Moon and find
\begin{align}
\frac{\hata_{\uparrow \fullmoon}}{\hata_{\uparrow \earth \fullmoon}}\simeq 2.9\times 10^{-3},
\end{align}
which is also small.

For an artificial satellite, we can estimate the order of the upper limit of its spin angular momentum using $J_m=Im_sR_s^2\omega$ and $\mathcal{O}(10^3~\text{kg}),\,\mathcal{O}(10~\text{m}),\,\mathcal{O}(100~\text{rps})$ for $m_s,\, R_s$ and $\omega$. A computation shows that this still yields an $\hata_m$ that is much smaller than the value of $\hata_M$ regardless of whether it orbits the Sun, any solar planet or the Moon in an equatorial orbit. This implies the effect of the self-spin of typical rounded satellites on their PS is usually smaller than the LT effect from the host objects.
However, if the satellite is specially designed (e.g. much wider separated masses rotating around the mass center) such that its specific angular momentum $a_m$
is larger than $3.0\times 10^6$ (m$^2$/s) if it orbits Mercury, or  $3.2\times 10^6$ (m$^2$/s) if it orbits the Moon, then the particle spin effect will be stronger than the LT effect from the host planet/moon. These values might be within the reach of current or near-future engineering. For satellites orbiting Earth or Sun, this will happen only if $a_m\gtrsim 1.2\times 10^9$ (m$^2$/s) or $a_m\gtrsim 9.6\times 10^{10}$ (m$^2$/s) respectively, which is probably too large to realize soon. In all these considerations, if the orbit inclination angle $\theta_{\mathrm{In}}$ is large, then the required $a_m$ can be reduced by a factor of $\cos\theta_{\mathrm{In}}$.

Finally, for stars orbiting supermassive BHs (SMBH), we can also estimate the $\hata_m$ by assuming that the star is either a fast spinning pulsar \cite{Hess2010} or a fast spinning normal star \cite{kath2020} up to ten solar masses. It is found that for SMBH with non-extremely small spin parameter $\hata_M$, including the Sgr A* which has $\hata_M\approx 0.9$ \cite{Daly:2023axh}, the effect to the PS due to $\hata_m$ will always be much smaller, by at least an order of $\mathcal{O}(10^{-4})$, than that of the LT effect of the central SMBH.

\subsection{Other spacetimes}

We also worked out the PS in other SAS spacetimes, including the Kerr-Sen \cite{Sen:1992ua}, Kerr-Taub-NUT \cite{KTN1966}, rotating
Ay\'on-Beato-Garcia \cite{Toshmatov:2014nya}, rotating Bardeen \cite{Bambi:2013ufa} and the rotating Hayward \cite{Abdujabbarov:2016hnw} spacetimes. However, as the asymptotic coefficients of the metric functions of these spacetimes agree with those of the Kerr ones up to the $p^{-3/2}$ order in the PS, their PS are indistinguishable from that of the Kerr spacetime up to this order. The unique parameter(s) in each spacetime only appears starting from the $p^{-2}$ order, which is currently not detectable and therefore we will not present the PS in these spacetimes here. In general, those PS are not difficult to obtain using Eq. \eqref{eq:psf} anyway.

\section{Discussions\label{sec:conc}}

In this letter, we computed the PS of spinning test particles in general SAS spacetimes using the PN method. The result is expressed as a half-integer power series of $(M/p)$ and the coefficients are polynomials of the asymptotic expansion coefficients of the metric functions, the particle spin and the eccentricity of the orbit. It is found that the particle spin can have a similar effect as the LT effect of the spacetime spin on the PS and both of them manifest linearly from the $p^{-3/2}$ order of the PS. The coupling between the spacetime and particle spins appears from the $p^{-2}$ order and has a sign similar to Ampere's force law between (anti)parallel current-carrying wires.
These results are then applied to the solar planets rotating around the Sun, the Moon around the Earth, satellites around the planets and the Moon and stars around SMBH. It is found that for Jupiter and Saturn rotating around the Sun, and specially designed satellites around Mercury or Moon, the self-spin of the orbiter can have a comparable effect on its PS as the spin from the central object.

There are a few directions that we can extend this work. The first is to study the precession along orbits not limited to the equatorial plane. The second is then to take into account the gravitational effect of the spinning particle on its motion. The latter is necessary if the particle mass is not negligibly small compared to the host and is particularly useful for studying gravitational radiations from their rotation.

\section*{Acknowledgements}

This work is supported partially by the NNSF China and MST China. Mr. Shaofei Xu would like to thank Mr. Letong Yang and Mr. Xudong Liao for their helpful discussions.

\end{document}